\documentclass[10pt,twocolumn,letterpaper]{article}

\usepackage[pagenumbers]{iccv} 

\usepackage{siunitx}
\usepackage{multirow}

\definecolor{iccvblue}{rgb}{0.21,0.49,0.74}
\usepackage[pagebackref,breaklinks,colorlinks,allcolors=iccvblue]{hyperref}

\def\paperID{*****} 
\def\confName{ICCV}
\def\confYear{2025}

\title{Resolution Revolution: A Physics-Guided Deep Learning Framework for Spatiotemporal Temperature Reconstruction}

\author{
Shengjie Liu$^1$ \quad Lu Zhang$^1$ \quad Siqin Wang$^1$\\
 $^1$University of Southern California\\
{\tt\small \{skrisliu\}@gmail.com,  \{lzhang63, siqinwan\}@usc.edu}
}

\begin{document}
\maketitle
\begin{abstract}
Central to Earth observation is the trade-off between spatial and temporal resolution. For temperature, this is especially critical because real-world applications require high spatiotemporal resolution data. Current technology allows for hourly temperature observations at 2 km, but only every 16 days at 100 m, a gap further exacerbated by cloud cover. Earth system models offer continuous hourly temperature data, but at a much coarser spatial resolution (9-31 km). Here, we present a physics-guided deep learning framework for temperature data reconstruction that integrates these two data sources. The proposed framework uses a convolutional neural network that incorporates the annual temperature cycle and includes a linear term to amplify the coarse Earth system model output into fine-scale temperature values observed from satellites. We evaluated this framework using data from two satellites, GOES-16 (2 km, hourly) and Landsat (100 m, every 16 days), and demonstrated effective temperature reconstruction with hold-out and in situ data across four datasets. This physics-guided deep learning framework opens new possibilities for generating high-resolution temperature data across spatial and temporal scales, under all weather conditions and globally.
\end{abstract}    
\section{Introduction}
\label{sec:intro}

Surface temperature is a critical physical property of the Earth's system and an important climate indicator~\cite{hansen2010global}. Over land, land surface temperature (LST) can be heterogeneous due to complex surface characteristics, with urbanization further exacerbating these differences~\cite{li2019evaluation}. Earth system models are developed through synthesizing satellite and meteorological station observations, using physics-based energy balance modeling to simulate surface dynamics~\cite{giorgi1997representation}. Currently, the best global simulations achieve a spatial resolution of 9~km over land areas and 31~km globally, much coarser than commonly used satellite data such as Landsat (100~m) and GOES-16 (2~km)~\cite{irons2012next}.

High-resolution temperature data are important for many real-world applications~\cite{liu2023spatial,li2023satellite}. However, the current lack of high spatiotemporal resolution temperature data has imposed limitations on many applications~\cite{wang2019environmental, wimberly2021satellite}. For example, central to causal inference in climate and health research is the use of time-series data to track changes in critical health statistics under varying daily temperatures. Yet, due to cloud cover and long revisit times, temperature data obtained from satellites are often incomplete, sometimes offering only 1--2 observations every 16 days~\cite{liu2025daily}. Cloud cover further exacerbates these limitations, with an average of 66\% of the Earth's surface obscured by clouds~\cite{king2013spatial}.
The lack of seamless daily temperature data has constrained the use of satellite-derived temperature products in real-world applications. As a result, many existing studies rely on temperature data from sparsely distributed meteorological stations, which offer limited spatial coverage and are unable to capture temperature gradients within cities~\cite{zhang2022hourly}.

Integrating coarse-resolution, continuous temperature data from Earth system models with high-resolution, cloud-contaminated satellite observations offers a promising pathway toward generating seamless daily temperature datasets~\cite{zhang2021practical, wu2021spatially, liu2025daily}. Despite the recent rise in high-resolution temperature data, the exponential growth in computational resources, and the rapid advancement of Earth system models, research in this area remains relatively limited, with existing studies adopting a variety of approaches from different perspectives.
For example, the reanalysis and thermal merging (RTM) method integrates 30~km reanalysis temperature data from a land data assimilation system and the 1~km MODIS observational data through multiple steps, using the annual temperature cycle (ATC) to capture the average trend, employing random forests to model daily fluctuations, and using a search-window algorithm to account for spatial residuals, achieving 2.03--3.98~K accuracy~\cite{zhang2021practical}. 
Similarly, a three-step approach was proposed to integrate 30~km reanalysis data and 4~km geostationary satellite data, with reconstruction accuracy at 3.57--3.94~K~\cite{ding2022reconstruction}. 
A more recent study proposed a two-step approach combining an enhanced ATC with ERA5 reanalysis temperature data and Gaussian processes to account for daily variation, achieving daily 30~m temperature data reconstruction at 1.48--2.11~K accuracy~\cite{liu2025daily}. 
These methods can be categorized as two-stage or multi-stage approaches, which separately capture the annual trend (often through the ATC) and daily fluctuations. However, they are typically hand-crafted and require a significant amount of manual processing and thresholding. End-to-end training has not yet been achieved in temperature reconstruction tasks.

\paragraph{Deep learning for temperature reconstruction.}
Deep learning models present an opportunity for a unified, end-to-end framework to reconstruct seamless surface temperature data~\cite{rasp2018deep, wegmann2023artificial, li2023climatenerf}. However, adapting deep learning approaches for satellite temperature reconstruction is challenging. A typical machine learning task requires pairs of training samples $\{x, y\}$, and the algorithm learns a mapping function $f$ such that
\begin{equation}
    y \sim f(x) \;,
\end{equation}
with $x$ as the input features and $y$ the target. Typically in Earth observation, $x$ is multispectral imagery, and $y$ is the corresponding label. For temperature data reconstruction, two factors limit its direct adaptation. First, temperature reconstruction involves time-series data that varies over time at a given location, yet the features $x$ are often only available monthly or even annually~\cite{li2023satellite}. Second, optical remote sensing features $x$ can only be collected when clouds are not present, but the primary interest often lies in reconstructing temperature under cloudy conditions. Another potential framework is super-resolution, where $x$ is coarse-resolution data and $y$ the corresponding high-resolution data~\cite{he2021spatial, lloyd2021optically}. However, integrating temporal dynamics remains difficult, and training still requires cloud-free pairs.

In this paper, we propose a different strategy: instead of directly predicting temperature, we predict the parameters of the ATC to enable full temperature reconstruction under all-weather conditions. We use the continuous coarse-resolution Earth system reanalysis temperature values as the signal and amplify them to capture daily fluctuations, and use convolutional layers (specifically U-Net) on features representing Earth surface properties to capture spatiotemporal dependencies and minimize bias. The proposed framework allows for unified, end-to-end training.

\section{Related Work}
\subsection{Image Inpainting: Spatial Interpolation}
In Earth observation, temperature data reconstruction is typically categorized into three distinct tasks. The first focuses on recovering missing values from a single time snapshot, where portions of a temperature image are obscured due to cloud cover or other events. This task is analogous to the \textit{image inpainting} vision task~\cite{yu2018generative}. This is the most-studied framework for temperature data reconstruction, with various methods having been developed and applied, including graph-based modeling, Gaussian processes, Fourier convolution techniques, and adversarial training~\cite{rolland2024improving, liu2024deep, bochow2025reconstructing, hirahara2019denoising, kadow2020artificial}. These approaches rely on the presence of at least some scattered observations at the same time within the spatial domain to guide the reconstruction. Unlike image inpainting, where visual plausibility is often sufficient and the plausible reconstruction solutions are acceptable, temperature reconstruction has hard ground truth, and it is essential to quantitatively evaluate the reconstruction results. 

\subsection{Two-Step Spatiotemporal Reconstruction}
Spatiotemporal reconstruction methods aim at overcoming the limitation from the same-time valid observations by interpolating through the temporal direction. Specific to Earth observation, repeated acquisitions of imagery over the same geographic area are common, making time-series information particularly valuable for temperature reconstruction. A typical spatiotemporal reconstruction approach applies a temporal reconstruction function, such as the ATC, to estimate temperature trends over time first, and then residuals between the reconstructed and observed values are computed and modeled using spatial reconstruction techniques, such as Gaussian processes, graph-based models, or window-based pixel similarity searches~\cite{liu2025daily, zhang2021global, zhu2022reconstruction}. These methods generally rely on hand-crafted features and involve separate steps for temporal and spatial modeling. As a result, the overall workflow is complicated and often requires manual intervention. Moreover, because temporal and spatial components are modeled independently, the final solution may be suboptimal due to the lack of joint optimization and end-to-end training.

\subsection{Change Detection and Super-Resolution}
An alternative framework, more directly aligned with modern deep learning paradigms, treats temperature reconstruction as a change detection problem. Given a fully observed thermal image and a subsequent image partially obscured by clouds, temperature changes can be computed over the observed pixels. With additional constraints, often derived from land cover information, these changes can be extrapolated to the obscured regions, enabling full-scene reconstruction~\cite{bahi2025new, long2020generation}. This approach can be further extended to super-resolution tasks~\cite{li2025lfsr}. Specifically, given a pair of aligned low-res and high-res thermal images, and a new low-res image, the goal is to generate the corresponding new high-res image. However, a common criticism of this framework is its reliance on clear-sky conditions~\cite{wu2021spatially, jia2024advances}. Under cloudy conditions, differences in surface energy balance result in temperature distributions that diverge significantly from those under clear skies. Consequently, change detection and super-resolution models trained on clear-sky data often fail to generalize, limiting their effectiveness in all-weather temperature reconstruction scenarios.

\section{Methodology}

\subsection{Overview: Physics-Guided Deep Learning With Time Consideration}
In the classical machine learning paradigm, given training samples \(\{x, y\}\), the goal is to learn a model \(\mathcal{M}\) such that
\begin{equation}
    y \sim \mathcal{M}(x).
\end{equation}
Typically, \(x\) is a set of features representing the Earth surface, which in vision tasks often correspond to images. This paradigm treats each sample independently, ignoring any temporal ordering. However, in temperature reconstruction tasks, time is a critical factor that must be explicitly incorporated:
\begin{equation}
    y(t) \sim \mathcal{M}(x, t).
\end{equation}
A na\"ive approach is to include time \(t\) as an additional input feature alongside \(x\). Yet, this does not exploit the intrinsic temporal structure of the data, and due to the flexibility of neural networks, it risks producing unrealistic predictions that deviate from physical reality.
To address this, we introduce the \emph{Annual Temperature Cycle} (ATC) as a physical constraint within the network architecture. The ATC component explicitly models the overall seasonal trend, capturing significant time-series characteristics. To account for daily fluctuations, we introduce one additional linear term applied to the coarse-resolution temperature data from ERA5. Finally, the convolutional layers focus on learning the remaining spatiotemporal variations only. Specifically, the proposed model comprises three additive components, all embedded within a CNN:
\begin{align}
    y(t, \mathbf{X}) \sim\ 
    &\mathcal{M}_{\mathrm{ATC}}\big(t \,\big|\, \phi_{\mathrm{ATC}}\big) 
    + \mathcal{M}_{\rho}\big(t \,\big|\, \phi_{\rho}\big) \notag \\
    &+ \mathcal{M}_{\mathrm{conv}}\big(\mathbf{X} \,\big|\, \phi_{\mathrm{conv}}\big),
\end{align}
where \(t\) is time,  \(\mathbf{X}\) is features representing the Earth surface, and \(\phi = \{\phi_{\mathrm{ATC}}, \phi_{\rho}, \phi_{\mathrm{conv}}\}\) denotes the learnable parameters of each component.

\subsection{Problem Setup: Temperature Reconstruction}
Let \(\mathbf{T} \in \mathbb{R}^{H \times W \times C}\) denote the input temperature tensor, where \(H\) and \(W\) are the height and width of the tensor, \(C\) is the number of time steps in the time-series, and some entries may be missing (NaN). We define a binary mask \(\mathbf{M} \in \{0,1\}^{H \times W \times C}\) indicating observed and missing values:
\begin{equation}
\mathbf{M}(i,j,t) = 
\begin{cases}
1, & \text{if } \mathbf{T}(i,j,t) \text{ is observed}, \\
0, & \text{if } \mathbf{T}(i,j,t) \text{ is missing (NaN)}.
\end{cases}
\end{equation}
We build a model \(g_\phi\) for temperature reconstruction, which is to map \(\mathbf{T}\) to its estimate \(\widehat{\mathbf{T}}\), composed of three additive modules:
\begin{align}
\widehat{\mathbf{T}} =\;
&\underbrace{\mathcal{M}_{\mathrm{ATC}}\big(\mathbf{T} \,\big|\, \phi_{\mathrm{ATC}}\big)}_{\text{annual temperature cycle}} 
+ \underbrace{\mathcal{M}_{\rho}\big(\mathbf{T}_c \,\big|\, \phi_{\rho}\big)}_{\text{daily fluctuation from ERA5}} \notag \\
&+ \underbrace{\mathcal{M}_{\mathrm{resid}}\big(\mathbf{X} \,\big|\, \phi_{\mathrm{resid}}\big)}_{\text{spatiotemporal correction}}.
\end{align}
where:
\begin{itemize}
    \item \(\mathcal{M}_{\mathrm{ATC}}: \mathbb{R}^{H \times W \times C} \rightarrow \mathbb{R}^{H \times W \times C}\) is a pixel-wise overall temperature trend over time, and we use the annual temperature cycle for modeling, with parameters \(\phi_{\mathrm{ATC}}\). Each pixel \((i,j)\) has its own learned parameters,
    \item \(\mathcal{M}_{\rho}: \mathbb{R}^{H \times W \times C} \rightarrow \mathbb{R}^{H \times W \times C}\) is a pixel-wise module that introduces daily temperature fluctuations. It applies a linear transformation to the resampled ERA5 temperature, denoted as \( \mathbf{T}_c \in \mathbb{R}^{H \times W \times C} \), and is parameterized by \( \phi_{\rho} \).
    \item \(\mathcal{M}_{\mathrm{resid}}: \mathbb{R}^{H \times W \times C_X} \rightarrow \mathbb{R}^{H \times W \times C}\) is a learned, high-frequency correction that accounts for unresolved spatial and temporal structure. This component replaces the spatial filtering process commonly used in two-stage approaches and is the key of the proposed framework to achieve end-to-end training in temperature reconstruction. 
\end{itemize}

\begin{figure*}[!t]
    \centering
    \includegraphics[width=0.92\linewidth]{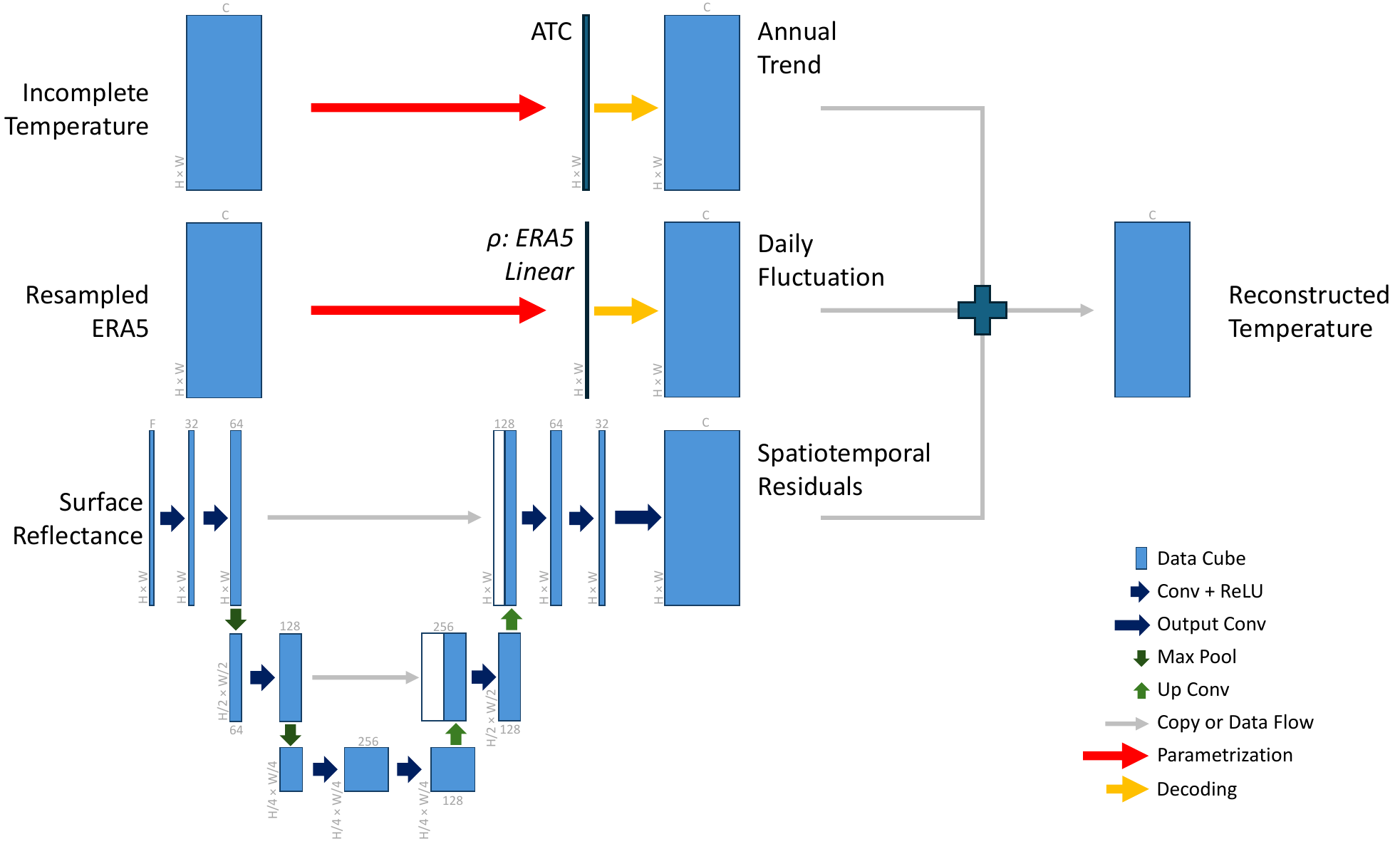}
    \caption{Overview of the proposed method. The annual trend and daily fluctuation are derived from physics-guided models (ATC and ERA5), while a U-Net-based CNN learns spatiotemporal residuals from surface reflectance. All components are combined in a unified, end-to-end trainable deep learning framework.}
    \label{fig:framework}
\end{figure*}

An overview of the proposed framework is shown in Fig.~\ref{fig:framework}. In the first branch, the incomplete temperature data are used to calculate the parameterization of the ATC model via optimization, and the ATC parameters are then decoded to reconstruct the annual trend. In the second branch, ERA5 reanalysis temperature data are used to parameterize a linear function, where the learned parameter $\rho$ amplifies the ERA5 temperature values to capture daily fluctuations. The third branch is a U-Net-based CNN that uses surface reflectance as input to construct any remaining spatiotemporal residuals not captured by the first two components. The three components are then combined to generate the final reconstructed temperature output. This is a unified, end-to-end training framework.

\subsection{Specific Model Components}
\paragraph{Annual Temperature Cycle.}  
The ATC component models the overall temporal temperature trend at each pixel \((i, j)\) using a cosine function:
\begin{equation}
\mathcal{M}_{\mathrm{ATC}}\big(i, j, t \,\big|\, \phi_{\mathrm{ATC}}\big) = a_{i,j} + b_{i,j} \cos\left( \frac{2\pi t}{T} + \varphi_{i,j} \right),
\end{equation}
where \(a_{i,j}\), \(b_{i,j}\), and \(\varphi_{i,j}\) are learnable parameters corresponding to the annual mean temperature, amplitude, and phase shift, respectively, and \(T\) is the period (e.g., 365 days).

\paragraph{Daily fluctuation from ERA5 Data with Partial Satellite Observations.}  
This component models the daily fluctuations using coarse-resolution Earth system model temperature data, denoted as \(\mathbf{T}_c(i, j, t)\), where \(\mathbf{T}_c \in \mathbb{R}^{H \times W \times T}\) represents the resampled data at the target resolution:
\begin{equation}
\mathcal{M}_{\rho}\big(i, j, t \,\big|\, \phi_{\rho}\big) = w_{i,j} \cdot \mathbf{T}_c(i, j, t),
\end{equation}
where \(w_{i,j}\) are pixel-wise learnable weights. The resampled temperature \(\mathbf{T}_c(i, j, t)\) is obtained via
\begin{equation}
\mathbf{T}_c(i, j, t) = \mathrm{Resample}\big(\widetilde{\mathbf{T}}_c(t)\big) \quad \text{at pixel } (i, j),
\end{equation}
where \(\widetilde{\mathbf{T}}_c(t)\) denotes the Earth system model temperature at its native resolution, and \(\mathrm{Resample}(\cdot)\) is an interpolation operator that maps the coarse-grid data to the fine grid indexed by \((i, j, t)\). In practice, the best available ERA5 temperature data has a spatial resolution of approximately 9~km over land, whereas the target resolutions in this study are 2~km and 100~m.

\paragraph{Spatiotemporal Correction Using U-Net-based CNN.}
The third component is a U-Net-based CNN designed to capture spatiotemporal residuals by modeling the discrepancy between the previous intermediate output and valid observations. The underlying assumption is that these residuals are associated with Earth surface properties, represented as \(\mathbf{X} \in \mathbb{R}^{H \times W \times C_X}\). The residual field is reconstructed using the following convolutional mapping:
\begin{equation}
\mathcal{M}_{\mathrm{resid}}\big(\mathbf{X} \mid \phi_{\mathrm{conv}}\big) = f_{\mathrm{CNN}}\big(\mathbf{X}; \phi_{\mathrm{conv}}\big),
\end{equation}
where \(f_{\mathrm{CNN}}\) denotes a CNN with a U-Net architecture, allowing effective multi-scale feature extraction and integration of both local details and global context. 

Specifically, we implement a simplified U-Net-style CNN to predict the remaining spatiotemporal residuals from spectral reflectance that are not captured by the previous two components. The U-Net architecture has a depth of three layers and takes the surface reflectance tensor \( \mathbf{X} \) as input. It transforms \( \mathbf{X} \) through the U-Net to reconstruct the temperature residual surfaces \( \mathbf{T}_{\text{resid}} \), which are then added to the outputs from the ATC (annual trend) and the ERA5 linear term (daily fluctuation) to generate the final temperature reconstruction.

\subsection{Loss Function}
To handle missing data, the reconstruction loss is computed only over observed values, using the mask \(\mathbf{M}\):
\begin{equation}
\mathcal{L}_{\mathrm{rec}}(\phi) = \frac{1}{\sum_{i,j,t} \mathbf{M}(i,j,t)} \left\| \mathbf{M} \odot \left( \widehat{\mathbf{T}} - \mathbf{T} \right) \right\|_1,
\end{equation}
where \(\odot\) denotes element-wise multiplication, and \(\|\cdot\|_1\) represents the element-wise L1 norm (i.e., the sum of absolute differences). We adopt the L1 loss instead of mean squared error (MSE) loss because it is less sensitive to outliers and typically yields better convergence during training.

\section{Results and Analysis}
We test the proposed framework and method on four datasets obtained from two different satellite sensors (Landsat, GOES-16), using hold-out and \textit{in situ} validation. We compare the proposed method with two exiting approaches: 4-step ATC~\cite{zhu2022reconstruction} and GEC-SEB~\cite{du2025reconstruction}. 

\subsection{Datasets and Experimental Setup}
\paragraph{Landsat Data}
Two separate datasets from the Landsat satellites are constructed. The raw thermal pixel obtained from the Landsat satellites is with 100~m ground sampling distance, and all data pixels are aligned with the multispectral sensor at 30~m on delivery. The Landsat surface temperature data is derived using a single channel algorithm, with 60\% of the observations within 2~K accuracy~\cite{laraby2018uncertainty}.
We obtained all data from 2023 over two separate locations where \textit{in situ} LST sites are available. The first dataset is over the PSU site ($40.7^\circ\mathrm{N},\ 77.9^\circ\mathrm{W}$) near State College, Pennsylvania. We obtain a total of 46 scenes within the 365-day period. After cloud-masking the data using the cloud masks provided along with the data, we subset a 256$\times$256 area centering at the PSU site. The second dataset is constructed over the BON site ($40.5^\circ\mathrm{N},\ 88.4^\circ\mathrm{W}$) near Bondville, Illinois. We obtain all scenes from 2023, resulting in a total of 91 available dates (within an overlapping area and thus doubled the frequency). We then again subset a 256$\times$256 area centering at the BON site. 

\paragraph{GOES-16 Geostationary Satellite Data}
Two separate datasets are constructed through the GOES-16 geostationary satellite. GOES-16, the first satellite in NOAA’s Geostationary Operational Environmental Satellites (GOES)-R series, was launched in November 2016 and enabled 2~km resolution temperature monitoring every 5 minutes for the first time~\cite{beale2019comparison}. The resulting LST data are produced hourly, with an accuracy of approximately 2.5~K when surface emissivity is known and proper atmospheric correction is applied, and around 5~K otherwise~\cite{Schmit2018ABI}. For the first dataset, we use all GOES-16 data from 2022 at 5:00 local time over a 700$\times$550 pixel region in the east coast centered on New York City, spanning from Qu\'ebec City to North Carolina and west to Detroit. The second dataset is another adjacent 700$\times$550 pixel region in the Midwest centering around Chicago and includes the Great Lakes at 19:00 local time. We use five GOES-16 spectral bands centered at 0.47, 0.64, 0.86, 1.61, and 2.24~\textmu m as input features. 

\paragraph{Validation Data}
Apart from the standard hold-out validation, we obtain the \textit{in situ} measurements from the Surface Radiation Budget (SURFRAD) network and calculate the \textit{in situ} LST value with the emissivity estimated using the ASTER dataset, following standard procedures~\cite{cheng2012estimating}. We used the annual mean spectral reflectance from the first seven spectral bands to represent the Earth's surface. The total number of available \textit{in situ} sites is limited (fewer than 10 in the U.S.), and in this study, each Landsat and GOES-16 dataset has one site within the coverage.

\paragraph{Experimental Setup} 
We conducted all experiments using PyTorch. The model was trained with the Adam optimizer at a learning rate of 0.1 for 500 epochs. For the Landsat datasets,  validation was performed exclusively using \textit{in situ} measurements due to the limited temporal coverage. For the GOES-16 dataset, we employed both \textit{in situ} validation and hold-out validation, with 20\% of the valid observed data held out for testing.

\subsection{Results on Landsat Data}

The results on the Landsat datasets are summarized in Table~\ref{tab:psu_method_compare} for the PSU site and in Table~\ref{tab:bon_method_compare} for the BON site. This evaluation uses \textit{in situ} measurements to assess all-weather LST reconstruction performance (a few days without valid \textit{in situ} measurements were excluded, resulting in 362 days for PSU and 359 days for BON in 2023).
At the PSU site, the proposed Physics-Guided CNN achieved the best overall performance (MAE = 2.75~K, RMSE = 3.44~K). Its bias was comparable to that of the 4-step ATC~\cite{zhu2022reconstruction}, an existing method, while achieving reconstructions across the full year (362 days), in contrast to only 48 days of the competitor. Another competing method published this year~\cite{du2025reconstruction}, based on lower-resolution inputs (1,000~m vs. 30~m), provided reconstruction for 229 days but yielded significantly inferior accuracy (RMSE = 4.93~K).
Similarly, at the BON site, the proposed method outperformed all competitors across all evaluation metrics, achieving an MAE of 3.43~K and a bias of 0.59~K. Overall, the proposed method enables accurate, daily, all-weather LST reconstruction at very high spatial resolution (30~m).

\begin{table}[htbp]
\centering
\caption{Comparison of Landsat temperature data reconstruction across different methods against \textit{in situ} measurements (PSU).}
\label{tab:psu_method_compare}
\scalebox{0.72}{
\begin{tabular}{lrrrrr}
\toprule
\textbf{Method} & \textbf{MAE (K)} & \textbf{RMSE (K)} & \textbf{Bias (K)} & \textbf{N} & \textbf{Resolution} \\
\midrule
4-step ATC~\cite{zhu2022reconstruction} & -- & 3.79 & \textbf{0.12} & 48$^{*}$  & 30~m \\
GEC-SEB~\cite{du2025reconstruction} & -- & 4.93 & -0.87 & 229$^{*}$ & 1,000~m  \\
\midrule
ATC                             & 3.97 & 5.19 & 0.66 & \textbf{362} & 30~m  \\
ATC + ERA5                      & 3.78 & 4.74 & 3.00 & \textbf{362} & 30~m  \\
Na\"ive CNN       & 14.63 & 17.68 &  -14.36 & \textbf{362} & 30~m  \\
\textbf{Proposed}             & \textbf{2.75} & \textbf{3.44} & -0.16 & \textbf{362} & 30~m  \\ 
\bottomrule
\addlinespace[2pt]
\multicolumn{6}{l}{\small\parbox{1.3\linewidth}{$^{*}$Existing methods are not daily for all-weather conditions and applied to selected days with observations only. For other days, existing methods reduce to ATC.}} \\
\end{tabular}
}
\end{table}

\begin{table}[!t]
\centering
\caption{Comparison of Landsat temperature data reconstruction across different methods against \textit{in situ} measurements (BON).}
\label{tab:bon_method_compare}
\scalebox{0.72}{
\begin{tabular}{lrrrrr}
\toprule
\textbf{Method} & \textbf{MAE (K)} & \textbf{RMSE (K)} & \textbf{Bias (K)} & \textbf{N} & \textbf{Resolution} \\
\midrule
4-step ATC~\cite{zhu2022reconstruction} & -- & 4.95 & 1.29 & 20$^{*}$  & 30~m \\
GEC-SEB~\cite{du2025reconstruction} & -- & 5.51 & 0.67 & 222$^{*}$ & 1,000~m  \\
\midrule
ATC                             & 4.51 & 5.76 & 2.85 & \textbf{359} & 30~m  \\
ATC + ERA5                      & 4.29 & 5.45 & 3.97 & \textbf{359} & 30~m  \\
Na\"ive CNN       & 15.45 & 19.15 & -14.93 & \textbf{359} & 30~m  \\
\textbf{Proposed}               & \textbf{3.43} & \textbf{4.42} & \textbf{0.59} & \textbf{359} & 30~m  \\ 
\bottomrule
\addlinespace[2pt]
\multicolumn{6}{l}{\small\parbox{1.3\linewidth}{$^{*}$Existing methods are not daily for all-weather conditions and applied to selected days with observations only. For other days, existing methods reduce to ATC.}} \\
\end{tabular}
}
\end{table}

\begin{figure*}[htbp]
    \centering
    \includegraphics[width=0.495\linewidth]{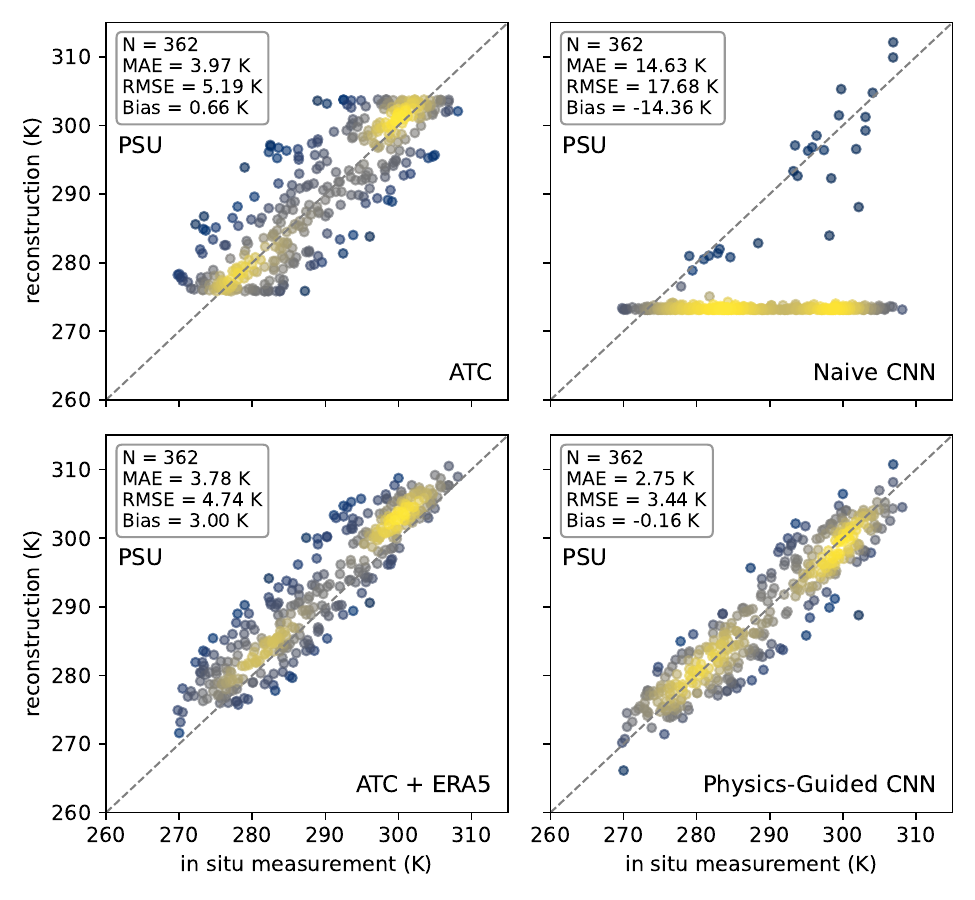}
    \includegraphics[width=0.495\linewidth]{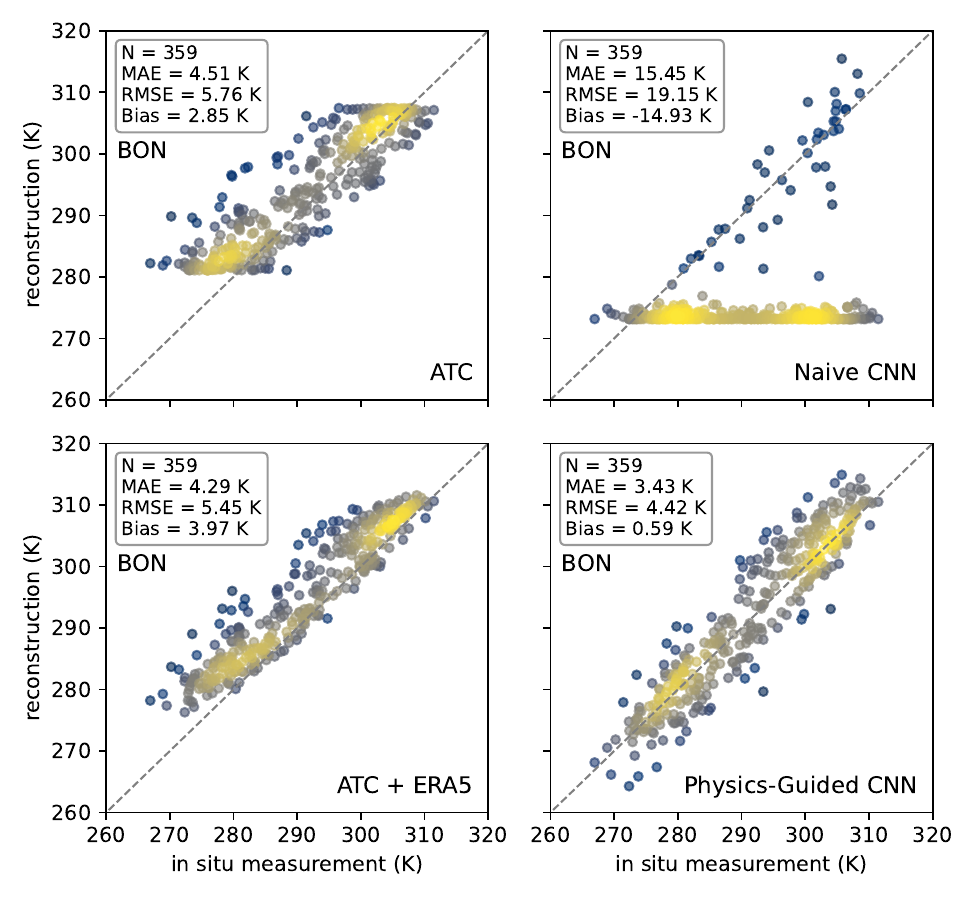}
    \caption{Comparison of reconstructed LST of Landsat (30~m) at PSU and BON sites from four models (ATC, Na\"ive CNN, ATC + ERA5, Physics-Guided CNN). Each subplot shows a density-colored scatter plot comparing \textit{in situ} measurements (x-axis) with reconstruction (y-axis).}
    \label{fig:plots}
\end{figure*}

\begin{figure*}[htbp]
    \centering
    \includegraphics[width=0.97\linewidth]{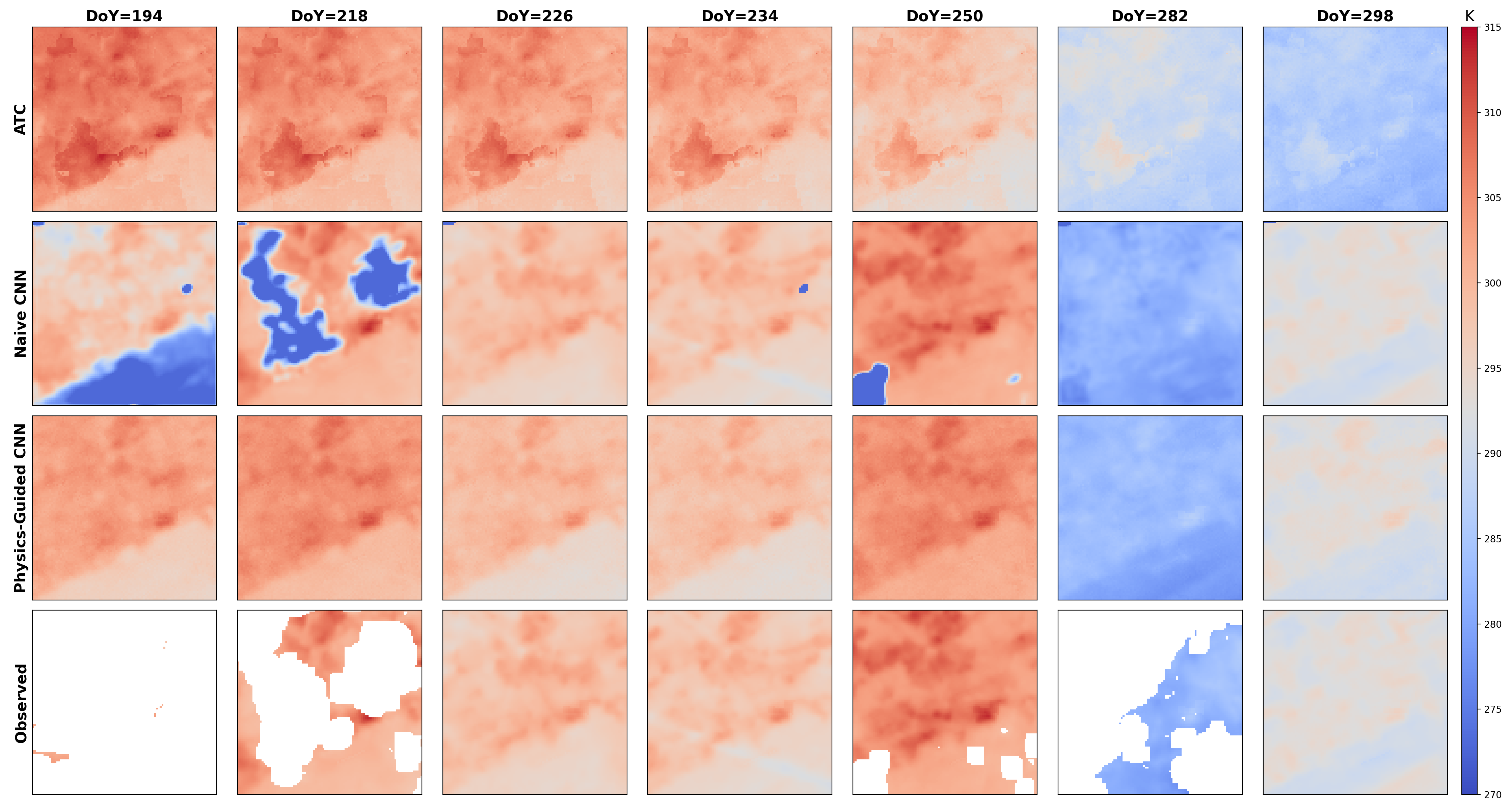}
    \caption{Reconstructed daily temperature maps at 30~m resolution from Landsat data.}
    \label{fig:psushow}
\end{figure*}

We show the scatter plots of the ATC, na\"ive CNN, ATC+ERA5, and the proposed Physics-Guided CNN at the two sites in Fig.~\ref{fig:plots}. The ATC reconstruction results are capped at a range of values, failing to capture temperature extremes at both the lower and higher ends. The na\"ive CNN method can only fit the data on dates with observations; for the remaining dates, the predictions show an extreme overfitting trend. The sparse data density over the temporal domain renders the na\"ive CNN ineffective in this case. The ATC+ERA5 method alleviates some issues related to capturing extremes, but a large bias still exists due to the lack of spatiotemporal correction. The proposed Physics-Guided CNN further reduces this bias by additionally accounting for spatiotemporal dependencies in the observational data. 

Finally, we show the reconstructed LST surfaces from seven selected days using ATC, na\"ive CNN, and the proposed Physics-Guided CNN, along with the valid partial observations, in Fig.~\ref{fig:psushow}. The reconstruction surfaces are biased to be too high (DoY: 194, 226, 234) or too low (DoY: 298) for the ATC model, due to its inability to capture daily fluctuations. Unrealistic reconstruction results are sometimes obtained (DoY: 194, 218, 250) using the na\"ive CNN due to its poor generalization ability. In contrast, the proposed Physics-Guided CNN consistently outperformed the competitors and achieved better results.

\subsection{Results on GOES-16 Data}
\begin{figure*}[!t]
    \centering
    \includegraphics[width=1.0\linewidth]{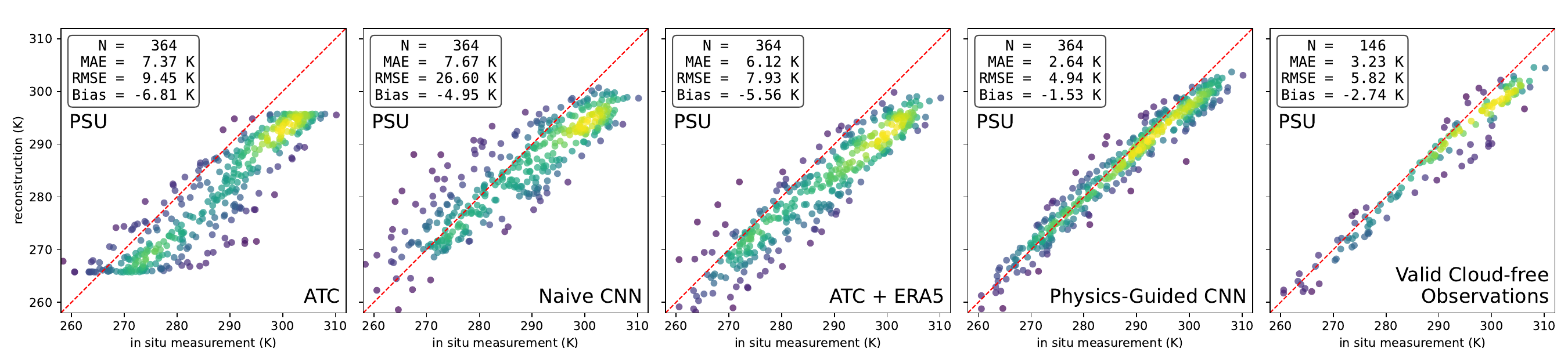}
    \includegraphics[width=1.0\linewidth]{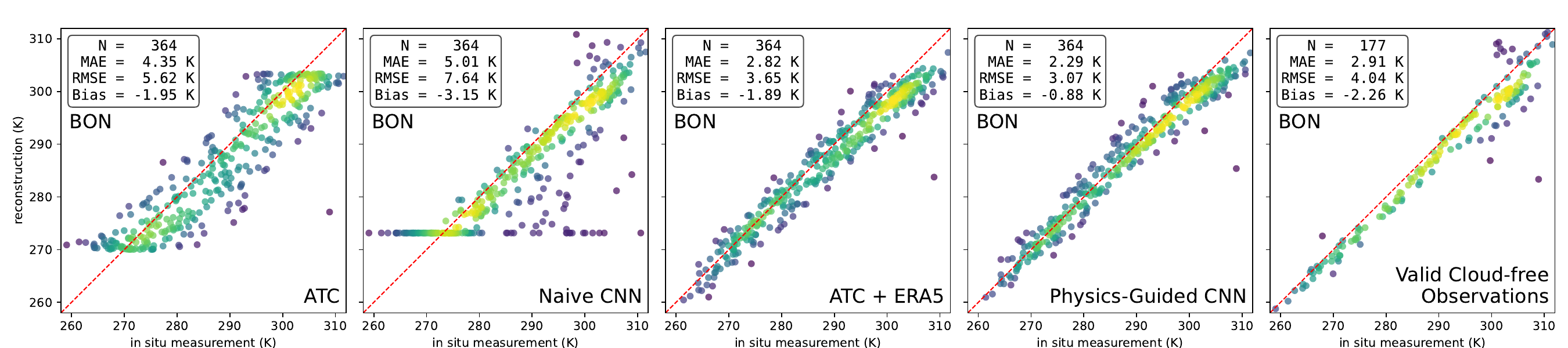}
    \caption{Comparison of GOES-16 LST reconstructions from ATC, Na\"ive CNN, ATC + ERA5, Physics-Guided CNN, and cloud-free observations at PSU and BON. Each subplot shows a density-colored scatter plot of \textit{in situ} (x-axis) vs. reconstructed values (y-axis).}
    \label{fig:goes16plot}
\end{figure*}

\begin{figure*}[!t]
    \centering
    \includegraphics[width=0.96\linewidth]{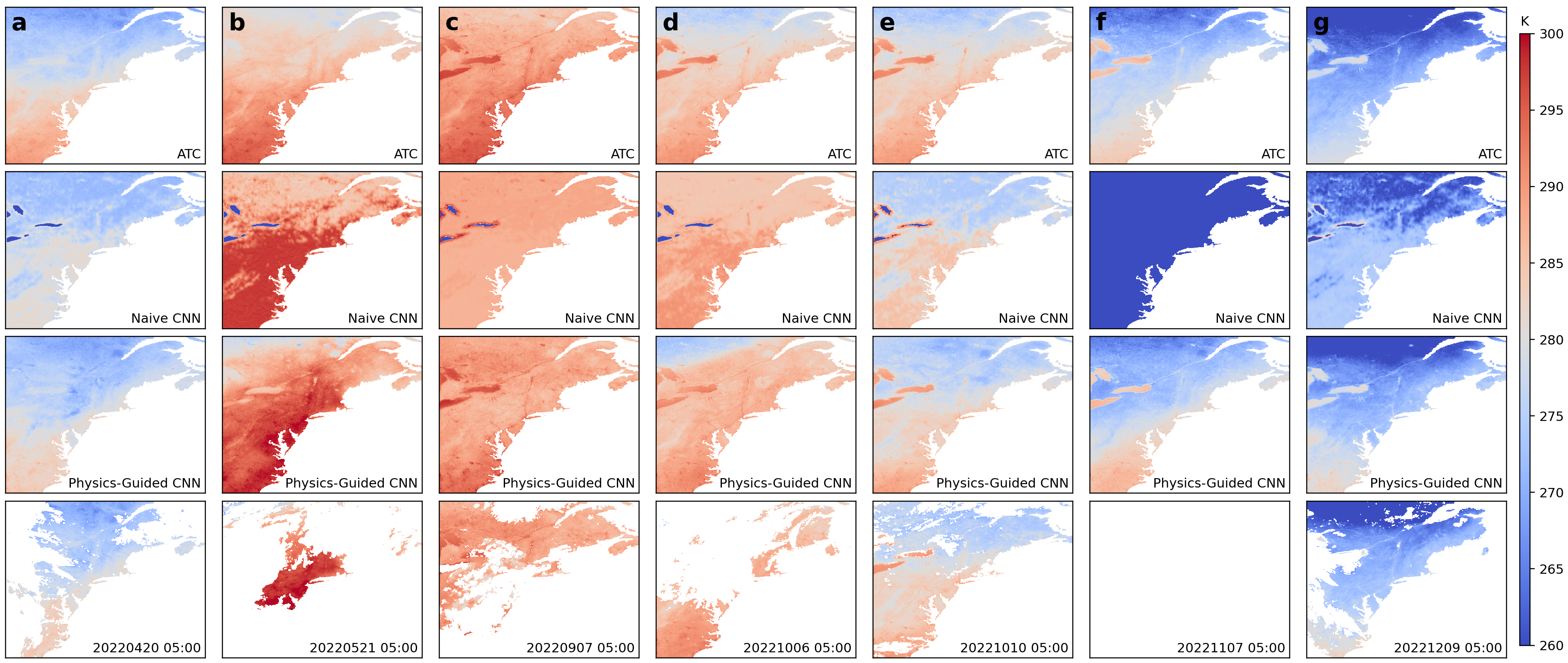}
    \caption{Reconstructed temperature maps of the GOES-16 data (East Coast including PSU). }
    \label{fig:GOES16results}
\end{figure*}

For the two datasets on GOES-16 data reconstruction, we first show \textit{in situ} validation through the scatter plots with the PSU and BON sites in Fig.~\ref{fig:goes16plot}. The observed pattern is similar to the Landsat reconstruction. For the PSU site, the proposed method achieving the best performance (RMSE = 3.72~K, N = 362). Notably, even the valid, cloud-free satellite observations (RMSE = 4.63~K, N = 145) yield worse results compared to the reconstruction. Similar pattern is observed over the BON site, with the Physics-Guided CNN achieving the best result (RMSE = 3.07~K) compared to the cloud-free observations (RMSE = 4.84~K), the ATC model (RMSE = 5.62~K), the Na\"ive CNN (RMSE = 7.64~K) and the ATC + ERA5 method (RMSE = 3.65~K).

The better agreement between the reconstruction results with the \textit{in situ} data compared to the valid-observed cloud-free data is likely due to better agreement between LST \textit{in situ} measurements and overhead observations under cloudy conditions~\cite{vancutsem2010evaluation}. Additionally, the imperfect match between the \textit{in situ} data and GOES-16 is likely caused by differences in field of view. The \textit{in situ} site has a field of view of approximately 70$\times$70~m, while the GOES-16 observations have a spatial resolution of about 2~km. Due to the averaging effect from coarser-resolution data, GOES-16 measurements tend to be lower than the \textit{in situ} values---a direction that is expected. Nonetheless, the results have demonstrated that the proposed method has achieved indistinguishable reconstruction compared with the observational data. 

We further evaluate the proposed method using the held-out 20\% test data, as shown in Table~\ref{tab:TableGOES16}. The proposed method achieves the best results among all, reducing the MAE metric by 52\% compared to the ATC method, from 3.75~K to 1.80~K. By observing the difference between training and testing evaluation metrics, we notice that the physics-guided approaches (ATC, ATC + ERA5, and the proposed method) tend to have good generalization ability, with comparable results between the training and testing data (e.g., for the proposed method, the MAE is 1.77~K for training and 1.80~K for testing). As we are optimizing the F1 loss (MAE), although the na\"ive CNN approach has a low loss, it fails to generalize and has very high RMSE values compared to other methods.

\begin{table}[!t]
  \centering
  \caption{Result comparison on the GOES-16 datasets.}
  \label{tab:TableGOES16}
  \scalebox{0.61}{
    \begin{tabular}{lrrrrrr}
      \toprule
      \multirow{2}{*}{\textbf{Method}} & \multicolumn{3}{c}{\textbf{Training Data}} & \multicolumn{3}{c}{\textbf{Test Data}} \\
      \cmidrule(lr){2-4} \cmidrule(lr){5-7}
      & \textbf{MAE (K)} & \textbf{RMSE (K)} & \textbf{Bias (K)} & \textbf{MAE (K)} & \textbf{RMSE (K)} & \textbf{Bias (K)} \\
      \midrule
      \multicolumn{7}{c}{\textbf{First GOES-16 Dataset (East Coast including PSU) }} \\
      \midrule
      ATC            & 3.63 & 4.91 & \textbf{0.05} & 3.75 & 5.02 & \textbf{0.05} \\
      Na\"ive CNN    & 3.67 & 15.85 & 0.59 & 3.68 & 15.91 & 0.60 \\
      ATC + ERA5     & 1.81 & 2.74 & -0.27 & 1.83 & 2.70 & -0.21 \\
      Proposed       & \textbf{1.77} & \textbf{2.64} & -0.18 & \textbf{1.80} & \textbf{2.67} & -0.18 \\
      \midrule
      \multicolumn{7}{c}{\textbf{Second GOES-16 Dataset (Midwest including BON)}} \\
      \midrule
      ATC            & 4.08 & 5.43 & \textbf{-0.16} & 4.21 & 5.55 & \textbf{-0.16} \\
      Na\"ive CNN    & 3.72 & 6.83 & -2.04 & 3.71 & 6.83 & -2.03 \\
      ATC + ERA5     & 2.12 & 2.98 & -0.34 & 2.09 & 2.82 & -0.27 \\
      Proposed       & \textbf{1.76} & \textbf{2.48} & -0.27 & \textbf{1.79} & \textbf{2.51} & -0.27 \\
      \bottomrule
    \end{tabular}
  }
\end{table}

To further demonstrate this difference, we show the reconstructed temperature maps from seven selected days, for the first GOES-16 dataset, in Fig.~\ref{fig:GOES16results}, covering a wide range of temperature values across all four seasons, and comparing the three methods (ATC, na\"ive CNN, Physics-Guided CNN). Among these, the proposed Physics-Guided CNN consistently demonstrates the best visual performance. For instance, in Fig.~\ref{fig:GOES16results}e, comparison with the near-complete valid observations reveals that the ATC model tends to overestimate temperatures in the northern region, while the na\"ive CNN produces unrealistic patterns, primarily due to its heavy reliance on spatiotemporal dependencies. In contrast, the Physics-Guided CNN generates reconstructions that most closely align with the valid observations.
On days without valid observations (e.g., Fig.~\ref{fig:GOES16results}f), the na\"ive CNN fails entirely to reconstruct the temperature field, as this becomes an extrapolation task—a scenario in which conventional deep learning models that rely heavily on statistical patterns often fall short. In contrast, physics-guided approaches, including both the ATC model and the Physics-Guided CNN, successfully produce plausible reconstructions. In other cases, the Physics-Guided CNN consistently outperforms the ATC model and other compositors.

\section{Conclusion}
In this study, we presented a physics-guided deep learning framework for land surface temperature data reconstruction, achieving, for the first time, end-to-end temperature reconstruction based on a single deep learning model. The proposed framework was tested on four datasets from two satellites with different resolutions (30~m daily for Landsat, 2~km hourly for GOES-16), achieving 2--3~K accuracy in hold-out validation for GOES-16, consistent with sensor’s accuracy, showing state-of-the-art performance on 30~m reconstruction for the Landsat datasets, and achieving better agreement with \textit{in situ} measurements on the GOES-16 data due to its ability to estimate cloud-covered temperature.
Central to the proposed framework is the encoding of a physics-guided model within a CNN and the usage of ERA5 reanalysis data to capture daily fluctuations. We used surface reflectance to reconstruct the residual surfaces. Recent Earth foundation models have shown promise to generate deep representations that better capture the Earth surface~\cite{marsocci2024pangaea, zhu2024foundations}. Integrating the deep representations can further enhance temperature reconstruction under the proposed framework.

{
    \small
    \bibliographystyle{ieeenat_fullname}
    \bibliography{main}
}

\end{document}